# Exploring and Eliciting Needs and Preferences from Editors for Wikidata Recommendations

Kholoud Alghamdi[1,2], Miaojing Shi[1] and Elena Simperl[1]

[1]Informatics, King's College London, London, United Kingdom.
[2]Data Science, University of Jeddah, Jeddah, Saudi Arabia.

[kholoud.alghamdi@kcl.ac.uk](kholoud.alghamdi@kcl.ac.uk); [miaojing.shi@kcl.ac.uk](miaojing.shi@kcl.ac.uk);
[elena.simperl@kcl.ac.uk](elena.simperl@kcl.ac.uk);

**Abstract**

Wikidata is an open knowledge graph created, managed, and maintained collaboratively by a global community of volunteers. As it continues to grow, it faces substantial editor engagement challenges, including acquiring new editors to tackle an increasing workload and retaining existing editors. Experiences from other online communities and peer-production systems, including Wikipedia, suggest that recommending tasks to editors could help with both. Our aim with this paper is to elicit the user requirements for a Wikidata recommendations system. We conduct a mixed-methods study with a thematic analysis of in-depth interviews with 31 Wikidata editors and three Wikimedia managers, complemented by a quantitative analysis of edit records of 3,740 Wikidata editors. The insights gained from the study help us outline design requirements for the Wikidata recommender system. We conclude with a discussion of the implications of this work and directions for future work.

# 1 Introduction

Wikidata is an open and collaborative knowledge graph (KG) launched by the Wikimedia Foundation in October 2012 (Vrandečić and Krötzsch, 2014).

It now has over 20,000 active editors, who have contributed to more than 90 million entities in the graph. While some tasks are undertaken automatically by bots, human editors remain at the core of Wikidata. They add content, keep it up to date, model knowledge as graph items and properties, and decide on content creation and management policies (Sarasua et al, 2019; Tharani, 2021).

As Wikidata continues to grow, it creates opportunities for editors to engage with a wide range of tasks. At the same time, such growth brings significant challenges in terms of community management (Sarasua et al, 2019). The sheer number of options available in Wikidata makes it difficult for volunteers to find relevant items to edit, especially when they are new to the community. Existing tools that automatically generate lists of tasks for editors to work on do not consider their interests, preferences, previous track record, or level of experience. Other support tools, such as QuickStatements,[1] which allows one to edit several items at once, or SPARQL,[2] which allows one to query the KG to find specific items, are only helpful once editors have identified productive ways to contribute. All this can impact engagement; logs made available openly by Wikidata show an uneven distribution, with the majority of the community performing only a few edits and a small number of active editors being responsible for almost 95% of all manual edits (Sarasua et al, 2019; Piscopo et al, 2017b; Cuong and Müller-Birn, 2016).

Experiences from other online communities and peer-production systems, for instance, Wikipedia, Quora, and the Zooniverse suggest that personalised task recommendations could help with retention and productivity (Cosley et al, 2007; Yang and Amatriain, 2016; Zaken et al, 2021), particularly when editors are relatively new to the community and must overcome so-called legitimate peripheral participation (LPP) effects to continue to engage (Piscopo et al, 2017b). Personalised recommendations could also be helpful for editors with more experience who sometimes struggle with a massive backlog of tasks to complete and need more nuanced strategies to prioritise this list (Moskalenko et al, 2020).

Our aim with this paper is to elicit the user requirements for a Wikidata recommendations system. To do so, we conduct an empirical study, following a mixed-methods approach that covers in-depth semi-structured interviews with Wikidata editors and the Wikimedia management team, supported by a quantitative analysis of edit logs of Wikidata item pages. The interviews with the editors (n=31) explore practices, needs, and preferences that are not easily observable or measurable through other means, in particular through the observational data that Wikidata makes publicly available. The interviews with the managers (n=3) elicit key considerations in recommender system design that are aligned with the volunteer community and the plans and ambitions of the core support team; specifically, we asked the managers for their opinions about the suitable presentation of the recommendations, the level of personalisation or diversity, as well as the level of control by the users. We use thematic analysis to identify recurring patterns in the responses and complement the findings with a descriptive statistical analysis of edit records of 3,740 randomly selected editors. The findings allow us to devise a list of requirements for a Wikidata recommender system.

---

[1] https://quickstatements.toolforge.org/
[2] https://query.wikidata.org/

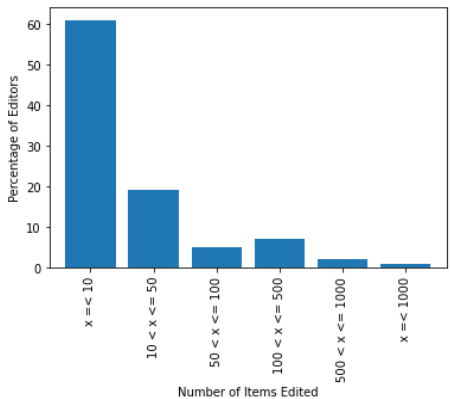

**Fig. 1**: Distribution of Edits per Editors

Our paper makes several key contributions: we provide insight into the needs, preferences, and practices of Wikidata editors, which can inform the design of a range of support tools; in our case, they help, alongside suggestions from the Wikidata management team and observational data from a sample of 3,740 editors, to devise requirements for a recommender system. This empirical study helps us identify fruitful directions of research at the intersection of HCI and machine learning.

## 2 Background and Related Work

### 2.1 Wikidata

Wikidata is an open, multilingual, freely available knowledge graph edited by a global community of volunteers from diverse backgrounds and experiences (Piscopo et al, 2017b; Geiger and Schader, 2014). There are currently over $300k$ registered Wikidata editors who can contribute to the KG with or without creating a user account. Like in other online communities and peer-production systems, participation is unevenly distributed (AlGhamdi et al, 2021; Piscopo et al, 2017b; Mu¨ller-Birn et al, 2015). As shown in Figure 1, around 60% of editors contribute to ten items or fewer; among them, there are newcomers who struggle to develop effective editing patterns in the long run and eventually drop out (Piscopo et al, 2017b; Sarasua et al, 2019). This is problematic as Wikidata continues to grow.

Wikidata stores data in the form of entities and organises it on pages. There are two types of entities: items and properties, located in different namespaces (Erxleben et al, 2014). Namespaces are identified by *URIs* using numbers and letters, with the letter *Q* representing items and the letter *P* relating to properties. Wikidata provides a form-based interface for editors to create and manage entities (Mu¨ller-Birn et al, 2015). Items and property pages have a similar interface, with language-specific labels, descriptions, aliases, and language-neutral statements (See Figure 2). Wikidata publishes a log of all edits of its entities. This is the data we focus on when implementing our recommender system. Other sources, such as talk and user pages, are less useful for our purpose. Many entities do not have any meaningful discussion data associated with them, and user pages are available for registered editors only and are very sparsely populated.

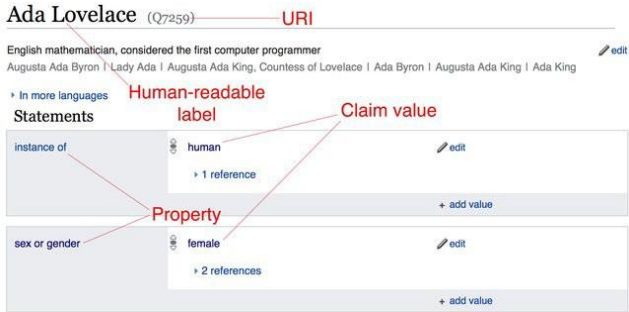

**Fig. 2**: The Structure of Wikidata Items (Piscopo and Simperl, 2018)

Edits in Wikidata are either manual, using an interface like in Figure 2, or automated, undertaken by running special scripts known as bots (Piscopo et al, 2017a). Manual edits can range from adding a label or a description in natural language to an existing entity to creating a new claim about an item to introducing new properties linking items. Some edits require greater insight into data and knowledge modelling and are not accessible to all editors; this is, for instance, the case with the ontological schema of the KG, which includes abstract classes of items as well as properties and property constraints (Piscopo and Simperl, 2018).

Previous studies have shown that as editors become more experienced, the volume and types of contributions evolve in tandem (Mu¨ller-Birn et al, 2015; Sarasua et al, 2019). Seasoned editors tend to use more tools, develop their own productivity practices, work on maintaining the structure of the KG rather than adding more entity data, and mentor other editors (Piscopo et al, 2017b; Sarasua et al, 2019). By contrast, newcomers utilise no or fewer tools and contribute to lightweight tasks, such as adding a new label or description to an item or property in their own languages (Piscopo et al, 2017b). In our paper, we complement prior studies by providing insights into the topical interests of editors and their evolution in time, which help us devise requirements for a Wikidata recommender system.

## 2.2  Human Factors in Designing Recommender Systems

Recommender systems (RS) provide users with personalised suggestions to help find relevant items of interest from a large collection of items (Aggarwal et al, 2016). Typically, designing a recommender system consists of four steps: (i) acquiring data about the interactions between users and items; (ii) creating users and items profiles; (iii) developing the recommendation algorithm; and (iv) evaluating it (Ricci et al, 2015). User profiles can be created in several ways, for example, by analysing users' interests and preferences based on the items they have explored, by considering users' demographics or interests, or by a combination of the two. Item profiles are based on attributes of the items which are relevant in the application domain. The recommendation algorithm predicts which items the user would prefer. Common strategies include: collaborative filtering (recommend based on similar user profiles), content-based filtering (recommend based on similar item profiles), and hybrid techniques (Aggarwal et al, 2016; Burke, 2002).

There is a large body of work considering human factors in recommender systems design, aiming to enhance user experience in multiple ways (Lex et al, 2021; Jugovac and Jannach, 2017). One set of studies incorporates user demographic attributes into the recommender model to improve the results. Demographic attributes are human features that have high stability and do not depend on a particular context. They hold implicit information about users' tastes and interests said2011comparison. Thereby, various systems (Tahmasebi et al, 2021; Al-Shamri and Bharadwaj, 2007; Al-Shamri, 2016) integrate demographic attributes when generating recommendations to alleviate cold-start issues (Tahmasebi et al, 2021), increase user satisfaction (Al-Shamri and Bharadwaj, 2007), and improve the level of personalisation (Al-Shamri, 2016). However, it is essential to mention that modelling such attributes needs to avoid ethical implications because this might disclose sensitive information about users (Al-Shamri, 2016). In a collaborative project like Wikidata, where users generally choose which items to work on, an individual's demographics may actually indicate their preferences within the project. Our study explores this aspect qualitatively by asking the editors whether any of their demographic attributes influence how they choose the topics in Wikidata. By doing this, we aim to identify whether it is useful to integrate the demographic features of editors into the recommender system or not.

The second set of studies in RS design models user behaviour by capturing long and short-term interests, as well as changes and temporal dynamics of these interests, to simulate human memory and decision-making processes. They suggest that people are more likely to trust a recommender system when it generates recommendations similar to their decision-making processes (Lex et al, 2021). There are various techniques to implement such models, e.g. recurrent neural networks (Liu et al, 2018), and time-based memory decay functions (Yang et al, 2019; Chmiel and Schubert, 2018). Our study explores the editing practices of the editors to understand if there is a sequential pattern in the editors' editing behaviour and if temporal factors play a role in editors' choices for the items they work on. We aim here to know how to effectively model the sequential behaviour of editors and temporal aspects in the recommender system.

The third set of studies aims to generate explainable recommendations to improve the transparency, persuasiveness, effectiveness, trustworthiness, and user satisfaction of recommendation systems. There are multiple types of recommendations' explanations in the literature, based on the kind of information source and recommendation algorithm used to generate them (Zhang and Chen, 2018). One of the most common styles relies on the items' features, in which the recommended item is explained based on its similarity with the items that the user has interacted with in the past (Papadimitriou et al, 2012). This approach can be found in content-based recommendation models, like the one proposed by (Vig et al, 2009), where movie tags are used as features to generate intuitive explanations for the recommendations. Collaborative filtering recommendation models can also utilise features of items to generate explainable recommendations, as proposed, for instance, in (Zhang et al, 2014). Further, an explainable recommender system can consider user demographics to tell the user why a certain item is more relevant to them (Zhang and Chen, 2018). In particular, an item recommended based on demographic information can be explained by, e.g., informing the user that "most users who live in your location bought this product". This type of explanation is found in (Zhao et al, 2016), which uses demographic information from social media accounts to generate descriptive explanations for product recommendations. User demographic features include but are not limited to age, gender, language, and residence location.

The quality of the generated explanations significantly influences how users judge and trust the recommendations, as argued by (Kunkel et al, 2019). They compare the quality of explanations curated manually with the quality of automatically generated explanations. Inspired by them, (Chang et al, 2016) propose to use crowdsourcing to create explanations. They first extract topical aspects of movies by collecting relevant review quotes for each aspect and asking crowd workers to synthesise the quotes into explanations. Then, they model users' preferences based on their activities and present explanations in a personalised fashion. Their user-based evaluation using 220 participants shows that human-based explanations yield better trust and satisfaction than personalised tag-based explanations.

## 2.3 Recommender Systems in Online Communities and Peer-production Platforms

In online communities and peer-production systems, there is a thread of works that propose using personalized recommendations for the crowds (AlGhamdi et al, 2021; Cosley et al, 2007; Moskalenko et al, 2020; Dror et al, 2011; Liu et al, 2017; Sun et al, 2018, 2019; Kurup and Sajeev, 2017; Safran and Che, 2018). Each of these works attempts to design an appropriate recommender system that can effectively deliver personalised and reliable recommendations by understanding the intrinsic characteristics of the application domain.

In Wikipedia, a first recommender system, SuggestBot, was proposed in (Cosley et al, 2007). It uses article titles, links, and co-editing patterns as main features to recommend articles to editors. A more recent recommender system by (Moskalenko et al, 2020) represents Wikipedia articles using Graph Convolutional Networks. Both works aim to recommend items (Wikipedia articles) to editors based on the features of items and editors. There is no explicit feedback that would confirm an editor's interest in an article. Equally, there is very little information about editors beyond what they have edited so far (Moskalenko et al, 2020).

In Wikidata, the recent model called WikidataRec was introduced in (AlGhamdi et al, 2021). It implements a neural network architecture that considers item contents, relations, and item-editor interactions. The task is formalised as a learning-to-rank problem where the learning function estimates the preference scores of editors to items so as to rank and recommend new items to editors. In specific, WikidataRec uses matrix factorisation to generate item-editor representations, sentence embeddings (ELMo) for item content representations, and TransR for item-relational representations based on the Wikidata knowledge graph. The end-to-end algorithm combines item-based representations via soft weights generated from a gating network. Then it computes the dot product between the merged item-based representation and the editor-based representation to predict preference scores for item recommendations.

In community question-answering (CQA), several papers develop recommender systems to route users to questions they might be interested in answering, hence improving their engagement on the platform and reducing question answering time (Dror et al, 2011; Liu et al, 2017; Sun et al, 2018, 2019). Their approaches follow specific design criteria to deal with the distinct features in the CQA. (Dror et al, 2011; Sun et al, 2018) model recommendation as a classification problem using different machine learning techniques. They are addressing the sparsity issue by applying various mechanisms: 1) implementing a hybrid approach with content and collaborative knowledge to exploit the different families of item descriptors; 2) capturing the diverse types of user-item interactions and differentiating between the types that are more indicative than others. (Liu et al, 2017; Sun et al, 2019) apply graph embedding techniques to tackle the same problem. Similarly, in Wikidata, sparsity issues exist where there are scarce interactions per user and item. Thereby, one of our design requirements is to follow a hybrid approach to mitigate the sparsity issues in the data.

Another related area is online crowdsourcing, where the aim is to allocate tasks published by a requester to an online crowd. (Kurup and Sajeev, 2017; Safran and Che, 2018) employ a probabilistic matrix factorization (PMF) to recommend suitable tasks to crowdworkers based on their previous activities, performance, and preferences. They handle the cold start problem by utilizing predefined categories (e.g., sentiment analysis, translation, image labelling) as additional features to improve the recommendation accuracy.

## 3 Methodology

### 3.1 Overview

We followed a mixed-methods approach, with an interview component including editors and managers from the core Wikimedia team and an observational component based on openly available edit logs. We ran interviews with editors to explore practices, needs, and preferences that are not directly measurable or observable (Kelly, 2009). We complemented these with interviews with managers to get direct insights and suggestions into the role of recommender capabilities in the Wikidata ecosystem and discuss core design parameters such as the level of personalisation and diversity of recommendations, the level of control editors may need, and the user interface. We analysed all interview data thematically. To expand on the findings, we also analysed a sample of edit records statistically.

By using this methodology, we wanted to combine the advantages of qualitative and quantitative methods (Bryman, 2006) to have a more complete picture of how a possible recommender system could look like. Our findings consider both in-depth, context-rich data from editors and the Wikidata team and system logs, which give us a way to observe and analyse editing behaviours in a granular and, at the same time, less intrusive way.

### 3.2 In-depth Interviews: Eliciting Editors' Practices, Needs and Preferences

We ran semi-structured interviews (average 20 min) with Wikidata editors via Zoom over four weeks in the summer of 2021. The semi-structured format meant that we could cover different aspects of participants' editing practices while also giving them the freedom to provide new, unexpected information. We asked editors to tell us about how they go about editing, about any tools and resources they may use to find items to edit, how they pick topics to work on and demographic factors that may influence their choices and practices.

Each interview had three parts (Appendix A). The first two parts were about participants' editing practices and their ways of choosing items. In the first part, we asked them to explain their initial editing experience when they joined Wikidata. The questions included how they picked what they wanted to work on, the edit tasks they preferred to perform, what topics they have worked on, whether these topics reflect their interests, and which tools they utilised to find items they wanted. The second part focused on their present experience and to which extent their practices have changed over time to help compare and identify changes in their editing methods and preferences. Finally, the third part enquired specifically about whether any of the editors' nationality, cultural background, specialism, and expertise influence their editing practices and the selection of t o p i c s .

All interviews were audio-recorded and subsequently transcribed. The interview study was approved by the local ethics committee of the authors' institution, with the registration confirmation reference number MRSP-20/21-23336. Participants consented to the interview and audio and video recording. All personally identifiable information was anonymized. Two researchers performed thematic analysis (Braun and Clarke, 2012) on the data using the Nvivo in two successive cycles. We applied two layers of coding in each cycle, similar to (Azungah, 2018). In the first layer, we used deductive categories: the topics that editors worked on at the beginning of their Wikidata tenure and now; how and which of their demographic attributes influences their choices for topics; what edit tasks they preferred to perform in Wikidata when they started and now; what tools they used when they were new and at present to find items they wanted to edit. Subsequently, we applied a second coding layer for each of these themes using an inductive approach (Campbell et al, 2013) to obtain further details. By running the analysis in two phases, we were able to look into the data at different levels of generality and from different viewpoints. We used the resulting themes to elicit the editors' practices and preferences during editing Wikidata as elaborated in the findings.

### *Recruitment*

We sought both new and more established editors. Newcomers are editors who have contributed to Wikidata for less than a year and are still exploring effective ways to participate. Established editors are those who have been active in Wikidata for more than a year and have edited several thousands of revisions. We recruited participants by posting a message on Wikidata's social media channels (Twitter, Facebook and Telegram) and by messaging them directly via their personal pages on Wikidata. In direct message recruitment, we were able to label newcomers and established editors by checking their registration date via the MediaWiki API [3], as well as their edit history records. From a total of 80 Wikidata editors invited, 31 eventually took part in the study. The sample included five females and 21 males, of which four were newcomers and 27 were established editors, according to our definitions. 28 of the participants were involved in other Wikimedia projects, mainly Wikipedia, before starting in Wikidata, while three joined Wikidata without prior participation in other Wikimedia projects. This sample accurately represents the Wikidata population in terms of both the gender and the tenure of the editors, according to the current proportion of females to males editors (which is 14% to 86%, as discussed in the papers (Klein et al, 2016; Zhang and Terveen, 2021)) and newcomers to established editors (which is 16% to 83.8%, based on a random subset taken from the Wikidata edits logs). Because we asked established editors to look back at their first experiences in Wikidata, this provided a generalisable qualitative insight about newcomers and mitigated their low number in our sample.

---

[3] https://www.wikidata.org/w/api.php

## 3.3 Logs Analysis: A Closer Look at Editor Behaviour

The interviews with editors provided us with insights into editing practices in Wikidata. To better understand and characterise these practices, we analysed a sample of Wikidata edit logs from several points of view: editors' topics diversity, editors' topics stability, the correlations between editors' tenure in Wikidata and their topics stability and their topics diversity. For the first aspect, we wanted to know if editors change topics within the same editing session. We used the same notion of edit session as in (Sarasua et al, 2019): a session is a time window in which an editor completes a series of (often related) edits, which are temporally isolated from other sequences of edits in their edit track record. The set of edits in the same session can be related to each other in terms of topics or in terms of editing actions. We used a threshold of 4 hours to define the length of an editing session (Sarasua et al, 2019). We define the one-topic session as a session where more than 60 to 80% of its edits belong to the same topic. Meanwhile, the multiple-topic session is the one where (¡60%) of its edits cover the same topic. We counted the number of one-topic sessions and multiple-topic sessions for each editor. We define one-topic editors as those whose most editing sessions (¿50%) are one-topic sessions. We define multiple-topic editors as those who are more than 50% of their editing sessions are multiple-topic sessions. For the second aspect, we aimed to explore whether the editors have short or long stability in their topical interests. Thereby, we looked at all topics someone edited during their tenure, and we calculated the overlap coefficient of topical interests for each editor, session-over-session ($SoS_T$) (i.e., the overlap of topics with the previous session). It is defined as below:

$$SoS_T = \frac{\mathsf{T}_{PS} \cap T_{CS}}{\mathsf{T}_{PS}} \qquad (1)$$

where, $T_{PS}$ is the set of topics in the previous session and $T_{CS}$ is the set of topics in the current session. After we have their intersection, we can easily compute the change in the editors' topics during their sessions as one minus the overlap coefficient (i.e., $SoS_{TC} = 1 - SoS_T$). Then, to calculate the overall topic stability of editors, we averaged $SoS_{TC}$ over all sessions and multiplied the final result by 100.

The topic is the generic domain that the Wikidata item belongs to. It is not readily available on the item's page, like the label and description. To this end, we assigned a topic to each item based on its description. We cast this task as multi-class classification, in which we trained a classifier that can categorise an item into a predefined topic. The classifier makes the assumption that each item is assigned to one and only one topic. We used the WDCM (Wikidata Concept Monitor) taxonomy of Wikidata items that encompasses 14 semantic categories [4] as a source of topics. The linearSVC classifier (Sahoo and Balaji, 2016) was utilised, and a training set of labelled data (i.e. the set of items-topics pair) was prepared and used to train the classifier. To prepare the training data, we followed the following steps: 1) we started with the list of topics in the WDCM and fetched the corresponding items of these topics (see Table 1). 2) using property path (?item wdt:P31/wdt:P279* ?class) [5], we extracted all items along with their descriptions that are instance of or sub-class of each item in the Table 1. The final training data was a list of items, their descriptions, and the classes representing their topics. This list was used as input to the classifier during the training process. The classifier was evaluated on the set of unseen data using Precision, Recall and F1-score, and the results were 0.87, 0.85 and 0.86, respectively.

For the third aspect, we wanted to understand whether the editors' topics' stability and diversity correlated with their tenure. To do so, we calculated the Pearson correlation coefficient (Benesty et al, 2009) of both the editor tenure and topics stability and editor tenure and topics diversity. To compute the editor tenure in Wikidata, we first fetched the registration date of each editor by querying MediaWiki API. We then calculated the number of days between each editor's registration day and the last day in our dataset.

We analysed edits performed on items' pages by human editors between 01.01.2019 and 01.06.2021. We obtained these activities from the XML data dump which is publicly available in Wikidata.[6] There are also editing activities performed on non-item pages, such as help pages or user pages. Our analysis looked at edits done on items alone because items are the main focus of our recommender - the aim to suggest items to edit rather than other types of community contributions. We distinguished between edits done using tools (such as QuickStatements, Wikidata Game, etc.) and without. To classify edits into these two groups, we scanned the edit comments for any digital traces left by tools listed in Wikimedia directories [7]. Then, we discarded all edits done using tools and kept only edits done without using tools. In the former case, editors do not decide what item or topic to work on.

Furthermore, we filtered out editors who edited fewer than five different items during our selected dates, which left us with a sample of 3,740 editors, 1,122,486 editing interactions with 1,109,695 items, which we used in our analysis.

The data has the following attributes: editor id, registration date, editor tenure, list of items the editor worked on, the topic of each item, and the time stamp of each edit.

---

[4] https://wikidata-analytics.wmcloud.org/app/WDCM$_S$ *emanticsDashboard*
[5] https://www.wikidata.org/wiki/Wikidata:SPARQL$_t$*utorial/en*
[6] Wikidata Wiki dump https://dumps.wikimedia.org/other/incr/wikidatawiki/
[7] https://hay.toolforge.org/directory/

**Table 1**: WDCM taxonomy with the corresponding Wikidata items.

| WDCM Taxonomy of ItemCategory | Wikidataitems Definitions | Wikidata i |
|---|---|---|
| Architectural Structure | All items that are related to buildings and other physical structures | Q811979 |
| Astronomical Object | All items that are related to planetary systems, galaxies ....etc | Q6999 |
| Book | All items that cover all kinds of books and novels | Q571 |
| Chemical Entities | All items that are related to Chemical Components and Elements | Q43460564 |
| Event | Items that cover History and Events | Q1656682 |
| Gene | All items that are related to Genes | Q7187 |
| Geographical Object | Items that cover Geography and Places | Q618123 |
| Human | Items that cover People and Self | Q5 |
| Organization | All items that cover all kinds of Organization (Companies, Universities.... etc) | Q43229 |
| Scientific Article | Items that cover all Scientific Articles in all subjects of Science | Q13442814 |
| Taxon | All items that are related to the organisms | Q16521 |
| Thoroughfare | All items that are related to transportation routes | Q83620 |
| Wikimedia | Items that cover all entities related to Wikimedia ecosystem | Q14204246 |
| Work of Art | Items that cover Art and Culture | Q838948 |

## 3.4 In-depth Interviews: Eliciting Wikidata Items Recommender System Design Requirements

We conducted interviews with two managers from Wikimedia Germany (the Wikimedia chapter leading on Wikidata) and one senior research officer from the Wikimedia Foundation in the fall of 2021. The participants were recruited via email by sending them a direct invitation to our interview study. The aim was to obtain further insights into editing behaviours and the broader Wikidata ecosystem to inform the design of a recommender system. We used semi-structured interviews with 15 minutes each, which were carried out via Zoom, audio-recorded and subsequently transcribed.

We focused on human factors that need to be considered in designing a recommender system for the Wikidata community, informed by existing literature in this space (see Section 2.2). We asked participants to provide their suggestions for designing a recommender system in terms of its interface design, the degree of diversity or personalisation generated by the system, and the required level of control by the editor. Again, we used thematic analysis to identify common themes in the answers.

# 4 Findings

## 4.1 Editors' Practices, Needs, and Preferences

We structure our findings around the themes used in our first layer of coding. For each theme, we report findings based on the thematic analysis of the interviews and the log analysis.

### 4.1.1 Editor Interviews and Log Analysis Findings 1: Topics and the Role of Demographic Attributes in Choosing Items

Most participants ($n = 28$) stated that their demographic attributes affect what topics they chose to different degrees and that these effects varied over time. As new editors, they started editing topics related to their *geographical locations*, *specialism*, and *subjects they love and have interests in*. The fact that these topics were either absent or short (needed more data) was an essential motivation for participants to edit.

- (P20) "I mainly started editing stuff I know for the area where I live, and most of these items weren't really filled."
- (P3) "Especially, in the beginning, people are often coming from a specific interest, that might be something you might call background, or it might be something like I have this hobby or something which is more this man's interested in something."

Some interviewees ($n = 6$) reported that, among their topics of interest, they tended mainly to edit the *human-related* items because, as (P9) justified, "It was easier to edit items that are people, which are just one entity." Furthermore, few participants ($n = 4$) mentioned that they started to edit *random* items, e.g.," everyday items" (P18), where they did not focus on any specific topics for the sake of "understanding how items were formatted" (P16) and "know more about data model of Wikidata" (P11).

Additionally, three participants stated that they sometimes tend to edit based on what others are editing:

- (P9)" I found some of my friends actually had items, so I added references for them."
- (P16) "I look at groups or things that people participated in."
- (P21) "I take a lot of inspiration from what the others are doing. I see that they're editing around similar things that I do."

As established editors, more than half of the participants reported that they still edit topics linked to their interests ($n = 8$), specialism ($n = 9$), and geographical location ($n = 11$). However, they stated that their topical preferences change over time as they engage with a different set of topics in each phase of their editing lifetime. In this context, we examined the stability of editors' topical preferences over time during editing records analysis. The calculations of $SoS_{T\,C}$ demonstrate that topical interests of editors change rapidly, with the average $SoS_{T\,C}$ being 58.14% (with standard deviation = 37.80%).

Moreover, we note that most of the interviewees ($n = 22$) tend to concentrate on a specific topic at a particular point in time, and then move to another topic in a sequential pattern. The following comments are indicative of the views of the participants on the matter:

- (P21) "I might have done a session where I did all those hospitals in Kazakhstan. When I'm done with that, I don't want to get recommendations for the next three weeks about hospitals or Kazakhstan because by then, it might be flowers in India or whatever."
- (P23) "Earlier this year, I created Wikidata items for all council seats that I've been in the Netherlands and the mayoral positions. Right now, I'm working on something entirely different."

**Table 2**: Counts for editors who tend to edit specific topics or several topics during their editing sessions

| # Editors who edited: specific topic and several topics (n=3,740) | |
|---|---|
| Specific topic in a session (1 topic) | 3,649 (97.83%) |
| Several topics in a session (>1 topic) | 81 (2.16%) |

The logs analysis confirmed this behaviour. As shown in Table 2, 3,649 editors of 3,740 editors (97.83%) were editing one topic in most sessions. By contrast, only 2.16% of editors edited more than one topic during most sessions.

Established editors also edit other topics outside of this original context, especially when those items need maintenance or have issues, such as lacking a label or a description, or not being linked to other items in Wikidata. Meanwhile, only a few participants ($n = 3$) focus entirely on solving issues and helping other members, without focusing on a particular topic. Returning to editors who started with random items, they stated that the scope of the tasks they perform becomes gradually more specialised. They currently edit specific topics related to their location or personal interests (P11, P16, P18, P30).

- (P16) "My topics have become more specific. I think I'm working on more things that are more historical, which is my interest currently."
- (P18) "So I have narrowed the scope. I'm working with things in my geographical space, like a geographical unit, and within that geographical unit, there are municipalities and districts, provinces, whatever."

In addition, interviewees' responses indicate that news-related topics play a role in their topic selection approach. Some participants ($n = 4$) stated that they tend to work on items about trending and breaking news, either local to them, such as "Dutch election" (P23), or global news such as "COVID-19" (P21).

Moreover, as explained in our methodology, we looked at the correlations between editors' tenure, topical preference stability, and topic diversity. The calculations of the Pearson correlation coefficient between editors' tenure and the stability of their topical interests illustrate that the relationship is not linear, with the value being −0.019. This is also illustrated in Figure 3, where we can see that the editors' lifetime in Wikidata is not correlated to their topics change percentage, meaning that both new and established editors have the same stability in their topical interests.

Again, we can notice no correlation between editors' tenure and their topics diversity (See Figure 4), in which the Pearson correlation coefficient is 0.448. This result suggests that all editors share similar practices regarding the number of topics edited in their sessions regardless of their tenure in Wikidata.

### *Summary of findings*

First, the interviews suggest that demographic attributes such as location, specialism and topical interests play a significant role in editing practices, specifically in selecting what items to work on. This is noted to be true for both newcomers and established editors. However, established editors also edit topics that are not directly related to their original interests for the sake of maintenance and keeping the contents up-to-date. Second, as noted by interviews and logs analysis, editors' topical interests change with time, and so they work on a particular topic at a time and then move to another topic in a sequence pattern. Third, from the interviews and logs analysis, we discovered a short stability of editors' topical preferences over time. Fourth, based on the log analysis, we can see no correlations between the editors' tenure and topical preference stability and topic diversity. Finally, a few participants in our interview pool mentioned specific strategies such as picking items randomly without concentrating on any specific topics and choosing topics related to the latest news or events. Figure 5 outlines the main results.

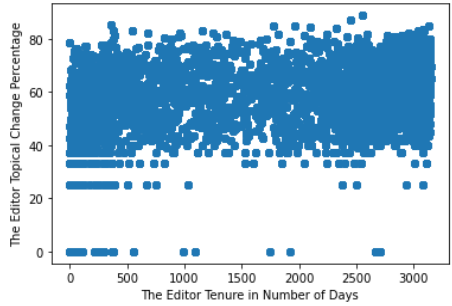

**Fig. 3**: The Editor Tenure vs. The Editor Topical Stability

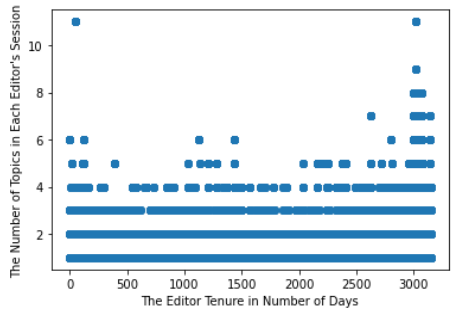

**Fig. 4**: The Editor Tenure vs. The Editor Topic Diversity

### 4.1.2 Editor Interviews Findings 2: Editing Tasks

We focus on item-related tasks in this study, including editing existing items and creating new items. The main results are outlined in Figure 6 which show that more than half of the participants ($n = 18$) started as newcomers by changing existing items rather than creating any items from scratch.

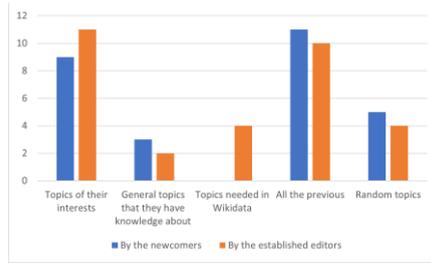

**Fig. 5**: The Topics that Editors Work on

One participant (P11) described this practice as "exploratory editing" that helps new members of the community better understand the editing experi-ence in Wikidata and learn more about the "kind of information [they] can represent" (P11). Additionally, it was easier for editors to start with these sorts of tasks, "without implementing anything in a wrong way" (P4), as at that time they "didn't know if rules were governing creating items" (P29). For these reasons, editing activities were mainly translation - adding missing labels and descriptions on items using their spoken languages. However, participants again emphasised that they were motivated to add more labels or descriptions because these were missing in Wikidata.

- (P27) "Not to create items in the beginning, but more to check items; add mainly Dutch versions of it. Translating data to Dutch, because most of it was in another language, but not in Dutch".

In line with their identity as former Wikipedians, some participants ($n = 9$) started with edits directly impacting Wikipedia, linking Wikipedia articles with Wikidata items by adding sitelinks. However, this behaviour seems more common among the interviewees who began using Wikidata in 2012 and 2013.

As editors gained experience, they became "less afraid to create new items" (P9) and more confident to work on items' statements sections (see Section 2.1), for instance, by adding properties to link items together within the Wikidata knowledge graph and enriching these items with internal and external resources. This confirms the findings in (Piscopo et al, 2017b), which showed how editors move from focusing on editing within a limited scope to taking on tasks that affect the KG as a whole.

- (P10) "I tend to enrich items manually, adding statements about the date of birth, place of birth, occupation, etc."
- (P12) "Trying to make sure things have references, citations added because I find that there's just so much uncited stuff on Wikidata very often."
- (P16) "Add properties to the empty items and connect items together (work from the bottom up)."

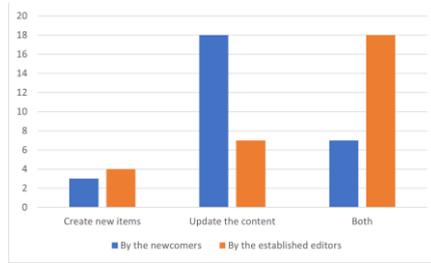

**Fig. 6**: The Editing Tasks that are Carried Out by the Editors

*Summary of findings*

Based on the interviews with editors, the common tasks noted among the newcomers are about editing existing items, such as translating the labels or descriptions of the items using editors' spoken languages. It is reported by interviewees that these types of tasks are more heuristic and hence, easier for novices. However, as editors become more experienced, they start to create new items and work on more advanced tasks, such as adding properties to connect items within the Wikidata knowledge graph.

### 4.1.3 Editor Interviews Findings 3: Tools Used by Editors to Find Wikidata Items

The most direct way to find items in Wikidata is through the search box, which was reportedly used by most interviewees ($n = 19$) when they started editing Wikidata (See Figure 7). One respondent stated: "In the beginning, I used the search box. I don't know if there are any other tools that can help decide which items to edit" (P7). The authors of (Piscopo et al, 2017b) discussed a similar behaviour, stating that newcomers tend to perform manual edits without utilising any tools. The search box is not an ideal way to find items. Another respondent stated: "I also used the basic search, but for most uses, it was not so helpful for finding groups of items needing intervention. It was not so efficient" (P10). These challenges in using the search box are due to its mechanism, which depends entirely on the similarity between the items users look for and the resulting pages (Wilson et al, 2010). Another participant stated, "it is dependent on which labels, descriptions, and aliases people put in for the data item; thus, it's sometimes hard to find the item you want, even if it already exists" (P29).

Additionally, some participants ($n = 3$) used Wikidata Game [8] to find the items to edit during their early days with Wikidata, which is a tool that presents editors with random items that need further work (e.g., missing an image, or a description, or requiring a link to Wikipedia articles).

---

[8] https://wikidata-game.toolforge.org/distributed/

- (P12) "I used Wiki Games or whatever, where it all presents random items missing a description or something like that. Most are fun to do a little bit."
- (P21) "The other thing that I've enjoyed is the Wikidata Game. I love playing that when I have nothing else to do."

This is in line with our earlier findings that some editors tend to start with broad and random topics. Furthermore, some participants (P8, P20) mentioned that, as newcomers, they used the links between entities to find new items of interest alongside the search box. In Wikidata, items that belong to similar topics are connected through a set of links (also called relations or properties).

Established editors appear to use the search box as a primary way to find items, in addition to more advanced tools. Several participants said that they started using the SPARQL query as a tool to find items only when they became aware of its existence in Wikidata.

- (P7) "When I became aware of the SPARQL query, then I started to use it to decide which items to edit."
- (P22) "Now, I use SPARQL queries as a mechanism for me to guide my work."

Our interviewees' responses suggest that the SPARQL query service is the most used tool to find items for editing in Wikidata. However, it needs editors to have adequate knowledge and skills to use it and sometimes does not provide a good user experience. One respondent (P9) stated: "It needs practice, and the data is not always good. So, you're running SPARQL queries on data that is messy. You have to know the query and how to refine it and refine your results to get what you want, which is like double the work." Another participant (P25) stressed that "I don't work with this query function, I try, but I fail, so I put it away again.". Thereby, as was pointed out in the introduction to this paper, the Wikidata recommendation system is developed as enhanced functionality that would be added to the existing tools in Wikidata (one option here is Wikidata Game) to generate personalised suggestions for the editors based on their edits history. Additionally, among other resources, some participants (P5, P10, P24, P11) stated that they tend to utilize Wikidata's weekly newsletter and the Current Highlights page[9] to find new items with missing content to edit them.

### *Summary of findings*

Based on the interviews, tools that editors use to find items in Wikidata are: 1) The search box, which is the most common tool newcomers and established editors tend to use while searching for items. However, it is noted that this tool is not efficient or helpful due to its difficulty in finding items containing non-descriptive data. 2) Wikidata Game, which presents editors with random items and simple edit requests. This tool is sometimes used by newcomers. 3) SPARQL query tool, which is an advanced tool mentioned by some participants who are experienced in Wikidata. However, besides the same limitations found in the search box, which are due to the underlying issues of the Wikidata data, the SPARQL query tool also requires a high level of knowledge from editors to be able to use it effectively. 4) The links that connect items together, Wikidata's weekly newsletter, and the Current Highlights section.

---

[9] https://www.wikidata.org/wiki/Wikidata:Main_Page

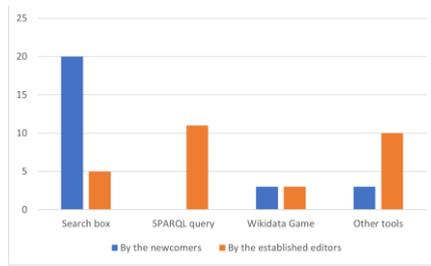

**Fig. 7**: The Tools Used by the Editors

## 4.2 Manager Interviews Findings: High-level Design Requirements Suggested by the Wikidata Team

Our three interviewees believed that presenting the recommended items as a long list is probably overwhelming, and editors will get bored. (P3) suggested having a ranked list of topics based on a matching or affinity score, where it can also be adjustable in a way editors can make it larger or smaller depending on their needs.

- (P3) "Maybe have an arrow at the bottom that they can click on to show them the fuller list. Perhaps the number five to eight to 10 and then maybe to increase it or expand it. Then, the editors can click on any of the recommended categories and see maybe more specific suggestions."

However, having a generalised design that fits every editor is impossible, as stated by (P2). (P1) also mentioned that "each set of editors have their own behaviour in editing the items, in which some are interested in picking a particular topic for the editing, yet, others want to do the same action, no matter which item is on."

When asked about the suitable level of diversity that the recommendations need to present, all interviewees agreed that having a diverse and balanced list of recommendations is advisable, especially for newcomers. Participant (P3) confirmed, "It would be beneficial for the new editors to have a personalised list, but this is something that you can give the user as an option of either seeing a personalised list or seeing a broader list of topics."

Giving editors control over the recommendations was perceived as essential by all participants. Based on suggestions provided by (P3), the control can be provided to editors in three approaches as follows: (1) Give them the option to control the length of the list, which going to be useful "if they are on a mobile device or just overwhelmed by a massive list of suggestions, they might want to filter it further." (2) Give them the option to set thresholds that allow them to control the level of diversity or personalisation of the results. (3) Give them the option "To say, don't ever show me results that pertain to this subject because knowing what people don't want to see can be just as useful as knowing what they want to see".

Transparency and explanation are additional design considerations that all the interviewees considered essential requirements to implement the recommender system. As noted in Section 2.2, explanation is one of the features that lead to a transparent system. The recommender system has to apply transparency because "the community wants to know why something is recommended to them and how it works." (P1). This can be done by "giving a lot of information about how you collected data and how the data have been used" (P3). Meanwhile, as two interviewees (P1, P3) stated, providing descriptive explanations for editors to understand why certain items or topics are recommended helps make the system more persuasive.

# 5 Designing a Wikidata Items Recommender System

Based on the previous findings, we now define the design requirements for the recommendation algorithm, features, techniques, evaluation metrics, and presentations of the recommender system (Table 3).

## 5.1 Personalised Profiles and Recommendation Algorithm

Interviews revealed that editors pick items based on multiple factors, related to their social and personal preferences, such as their topics of interest, specialism, location, and the editing activities of other editors. Therefore, the recommender system should apply a hybrid of collaborative and content-based techniques to model the editors' preferences and practices effectively. Here, the topical interests of editors could be obtained either by explicitly asking them to provide information about their interests (for instance, via a questionnaire) or from their past editing activities, specifically items they have edited. The editors' specialism and location could be collected by directly asking them. Meanwhile, the model can identify similar editors based on their editing affinities by applying collaborative-filtering techniques.

Moreover, we learned from the interviews that Wikidata editors are not a homogeneous group, in which the preferences of each individual are different from the others. Thereby, the recommender system should be able to model each editor's preferences effectively by capturing informative behaviours from the previous editing activities and assigning them with more weights.

## 5.2 Context Awareness

The editors tend to edit a specific topic and then move to another topic according to their contexts and preferences. However, it is also clear from the logs analysis that there is only short-term stability in topical interests for most editors. This may imply that topics of their most recent editing activities are more important than their long-term preferences. Therefore, the recommender system should capture these sequential dependencies and temporal dynamics effectively. Interviews also revealed other indicative contextual signals for editors' preferences, beyond the sequential editing behaviour, including location and the time-based popularity of items (such as COVID19-related or elections-related items). Thus, the recommender system should be able to formulate recommendations considering different kinds of contextual factors, and that will likely take effects, such as editors' locations and items' temporal popularity.

## 5.3 Degree of Personalisation

The interviews showed that editors sometimes focus on editing a set of topics that belong to their interests. Other times, they prefer to edit without any focus on a particular topic. Here, diversification in recommendations helps to increase the probability of retrieving novel items and achieving the recommender system's underlying goal of introducing editors with new items. This can raise the quality of the editor's experience with the system (Kunaver and Požrl, 2017; Helberger et al, 2018). However, it needs to be achieved while keeping high accuracy to match the findings of our    analysis.

## 5.4 Recommendations Presentation

Although the Wikidata recommender is not proposed to be a standalone tool with a user interface but a functionality that will be added on top of existing Wikidata tools, we can still take insights from the interviews in terms of the recommendations' presentation. First, we learned from the interviews that editors in the items selection process tend to pick items depending on topics of interest or topics that belong to their domains of knowledge or locations. Thereby, the recommender system shall display the recommended items as lists grouped by topics. In this context, some literature suggests that grouping the items into meaningful clusters can make the decision process easier for users and presumably increase their satisfaction (Jugovac and Jannach, 2017; Nanou et al, 2010; Chen and Pu, 2008). Furthermore, as suggested by the Wikimedia team, topics should be ranked based on their relevance to the editors, and items in each group should also be ordered according to their assumed relevance to the editors.

Finally, in their interviews, the Wikimedia team affirmed that editors need to understand the underlying recommendations strategy on generating the suggestions because this makes them trust the recommender. Thus, the recommender system should provide meaningful list headers, such as "Inspired  by your editing history, the following items are suggested for you," or "The other editors who edited similar items to you also edit this collection of items.". Such headers help provide explanatory characters for the recommendations (Jugovac and Jannach, 2017).

## 5.5  User Control

From the interviews, we acknowledge that it is not easy for individuals to describe their needs and preferences which are usually multifaceted and change over time. Therefore, the recommender system should represent information in ways that empower editors, giving them the control over their tasks and information needs. This was considered essential by the Wikidata team, who know the community well. User control can be realised in various ways, as proposed by the Wikimedia team: (i) by enabling editors to have the control to fine-tune the desired level of accuracy and diversity, (ii) by giving editors with means to re-rank recommendation lists and control their length, and
(iii) by allowing editors to express their explicit disinterest with generated recommendations.

Allowing the users to control the system to meet their personal needs has proven to increase their satisfaction and trust with the recommender system (Vaccaro et al, 2018; Jugovac and Jannach, 2017). A study by (Schafer et al, 2002) shows that users prefer to control the display of the data, in which they tend to view a larger recommendation set when they are able to filter it by themselves. This suggests that they already appreciate this simple type of control.

## 5.6  Evaluation: Measuring Effectiveness

Since one of the design requirements is to generate the recommendations as lists of ranked items grouped by topic (see Section 5.4), the recommender system's performance should be optimised and evaluated for the relevance of produced topics to editors and the accurate ranking of items. Thereby, precision, recall, F1 metrics, and list-wise metrics such as mean average recall (MAR), mean average precision (MAP), Mean Reciprocal Rank (MRR), and Normalized Discounted Cumulative Gain (NDCG), are the most appropriate measures for assessing the quality of recommended topics and items (Cremonesi et al, 2010).

Beyond the statistically measured quality of the recommender system, criteria related to editors' experience should also be evaluated. The results of an empirical study by (Cremonesi et al, 2012), reveal that user's judgment towards an RS's quality is likely to be more affected by factors related to the user experience. Hence, in our case, to accommodate editors' practices in editing diverse items revealed by the interviews and logs analysis, the recommender system should be optimised and evaluated on its ability to diversify the recommendations. Based on the findings of our editors' study, this criterion is expected to lead to more satisfying recommendation lists for editors. Moreover, it is also important to conduct user-centric evaluation, as (Pu et al, 2011) and (Knijnenburg et al, 2011) (i.e., online evaluation) to validate the claim that the recommender system improves an editor's experience.

**Table 3**: Summary of Requirements for the Wikidata Items Recommender System Design and How they are Implemented in WikidataRec

| Requirements / The Implementation in WikidataRec | Descriptions | The Source of the Elicitation |
|---|---|---|
| Personalised profiles and recommendation algorithm | • The recommender system should follow a hybrid approach combining collaborative and content-based techniques to effectively model editors' preferences and practices.<br>• The recommender system should model editor's preferences effectively by capturing informative behaviours from the previous activities and assigning them with more weights. | Editor Interviews Findings 1 |
| Context awareness | • As editing behaviours of editors have indicated short-term stability in their preferences, the recommender system should capture all the sequential dependencies that distinguish between older and more recent interests.<br>• The recommender system should have a high level of context awareness and use of real-time information, such as editors' locations and items' temporal popularity. | Editor Logs Analysis Findings 1 |
| Degree of personalisation | • The recommender system should use mechanisms that can generate diverse recommendations while keeping high accuracy to cope with the editors' practices' variety. | Editor Interviews Findings 1 and Editor Interviews Findings 3 |
| Recommendations Presentation | • The recommender system shall display the recommended items as multiple lists grouped by topics.<br>• In the results list, topics should be ranked based on their relevance to editors, and items in each group should also be ordered according to their assumed relevance to an editor.<br>• The recommender system should provide explanatory and meaningful list headers that aid editors to understand the underlying recommendations strategy of the results. | Editor Interviews Findings 1 and Manager Interviews Findings |
| User control | • The recommender system should represent information in ways that empower editors, giving them the control over their tasks and information needs, as follows:<br>1. By enabling the editors to have the control to fine-tune the desired level of accuracy and diversity.<br>2. By giving the editors with means to re-rank the recommendation lists and control their length.<br>3. By allowing the editors to express their explicit disinterest with the generated recommendations. | Manager Interviews Findings |
| Evaluation: measuring effectiveness | • The recommender system's performance should be optimised and evaluated for both the relevance of the produced topics to editors and the accurate ranking of items.<br>• Common quantitative measures and list-wise metrics such as precision, recall, F1, Mean Reciprocal Rank (MRR), and Normalized Discounted Cumulative Gain (NDCG) should be utilised to assess the quality of recommended topics and items.<br>• The recommender system should also be optimised and evaluated on its ability to diversify recommendations.<br>• The recommender system should be evaluated using user-centric evaluation to validate the claim that the recommender system improves an editor's experience. | Editor Interviews Findings 1 |

# 6 Discussion and Directions for Future Research

Understanding editors' practices and preferences in editing Wikidata, as we have done in Sections 3.2, 3.3, and 3.4 plays an important role in designing a Wikidata item recommender system. For example, the most recent system, WikidataRec (AlGhamdi et al, 2021), implements some of the design requirements. The system is evaluated offline using only historical edit data rather than user feedback, with promising results, which demonstrate that the design can produce accurate recommendations.

In the following, we discuss the system only in the context of the design requirements from Section 5 and suggest future directions of work influenced by the findings of the present study.

## 6.1 Implementing the Design Requirements

### 6.1.1 Personalised Profiles and Recommendation Algorithm

WikidataRec uses a hybrid recommender strategy that combines content-based and collaborative filtering techniques to rank items for editors. The approach effectively considers the two significant factors that are revealed by editors' interviews: editors select items depending on their topical interests and the items that other editors have edited. The technical implementation exploits both editors' previous interactions with items and items' content and relations to derive the recommendations. Furthermore, the approach uses gating net-work techniques to tackle the heterogeneous editors' preferences by assigning different weights on item-based representations depending on inputs of editors and items.

We learned earlier that editors' specialism and location play a role in their editing preferences. It would be interesting to further incorporate them into the Wikidata recommender system in future to evaluate the benefits, especially for newcomers who do not have previous editing activities. The editors' locations can be integrated using the *Nearby* [10] feature that is available in Wikidata, which asks the editors to provide their GPS data to generate tailored recommendations. Meanwhile, editors' specialism can be incorporated by asking the editors explicitly to provide this information (e.g. through using an initial questionnaire).

### 6.1.2 Recommendations Presentation

WikidataRec is proposed in (AlGhamdi et al, 2021) as a machine learning model, which has not been implemented in Wikidata yet. However, we can still take insights from its results in terms of the recommendations' presentation. In this, we noted that WikidataRec assigns different weights to different item-based representations. This is based on the observation that different item-related representations contribute non-equally to the final recommendations. Hence, the soft gating approach could be helpful in generating explanations for editors to justify why the algorithm recommends certain items. This aligns with one of the design requirements described in Section 5.4. In this context, future research should focus on developing a recommender model that can generate persuasive explanations (while keeping accurate recommendations) by utilizing user-based and item-based sources. Accurate recommendations accompanied by persuasive explanations help obtain the community's trust for recommendations (Gedikli et al, 2014).

---

[10] https://www.wikidata.org/wiki/Special:Nearby/coord/50.9280256,-1.3991936

### 6.1.3 Context Awareness

WikidataRec (AlGhamdi et al, 2021) has showed promising results in the offline evaluation. Notwithstanding, it did not model the sequential dependecies among editors' interactions. A promising direction derived from the design requirements in Section 5.4 is to apply sequential learning techniques that can distinguish between older and more recent editor interests and to take contextual information such as items' temporal aspects and popularity into recommendations' account.

## 7 Conclusion and Future Works

To this day, only a handful of tools exist in Wikidata to help editors find items of interest. These tools do not take into account editors' interests or preferences, and our study suggests that they have several other limitations. Our aim was to elicit the user requirements for a Wikidata recommendations system in the broader context of Wikidata tools and editing practices.

To this end, we ran an empirical mixed-methods study covering in-depth interviews with 31 (newer and more established) editors and three members of the Wikimedia management team and analysed the edit records of 3,740 editors statistically. The study helped us understand how editors choose items to edit, which tools they utilise, which editing tasks they tend to perform, how they pick topics to work on, and whether their demographic characteristics influence their choices and practices. Additionally, we gathered valuable suggestions from managers about high-level design considerations for Wikidata items recommender system.

Still, our study has some limitations worth mentioning. First, regarding the analysis of edit records, while we conducted it to expand our interviews' findings by including a second source of data based on a much larger sample, it covers only a specific time window of editors' edits. However, we were able to get a rich understanding of peoples' interactions with Wikidata items. Second, the interviews with the Wikimedia management team focused on obtaining suggestions for the user-centred design of the recommender system. These suggestions have not been implemented yet in the recent system (WikidataRec). However, since the aim of the paper is to open the path towards developing Wikidata items recommender system hence, the design suggestions we collected from the management team are useful for future studies that will build their decisions based upon the findings of this paper. Third, we investigated the role of editors' demographic attributes in choosing topics to work on. Yet, we assume that the enduring personality traits of editors might also influence their editing practices. Thus, additional research is needed to investigate this aspect. It would be interesting to examine whether there is any link between individuals' personality traits and their editing behaviours and study the feasibility of incorporating such information into the recommender system.

## Declarations

- **Funding** The authors did not receive support from any organization for the submitted work.
- **Ethics approval** The interview study was approved by the local ethics committee of the authors' institution, with the registration confirmation reference number MRSP-20/21-23336.
- **Consent to participate** Participants consented to the interview and audio and video recording.

## Appendix A   Interview Questions

1. What is your Wikidata username?
2. How long have you contributed to Wikidata?
3. When you started editing in Wikidata for the first time, tell me about your experience in deciding what items to edit?
4. What topics did you work on? Please specify one or more from the following options:
   - you worked on topics that you were interested in
   - you worked on general topics you had knowledge about
   - you worked on topics that you thought were needed in Wikidata
5. Which of the tools available by Wikidata were you using when you started? Did you use for instance:
   - Search box
   - Open task page
   - Specific tools (please identify)
   - Other ways (please specify)
6. Which editing task did you prefer doing when you started? Please choose:
   - Create new items
   - Update the content of the existing items
   - All of the above options
7. At the moment, how do you choose what to work on (items for the editing)?
8. What has changed in practice in terms of topics you choose to edit and tools you use, and the editing task you tend to perform?
9. Do you think any of your demographic attributes like nationality, cultural background, specialism, and expertise influence your strategy to choose what to work on?